# Light pollution offshore: zenithal sky glow measurements in the Mediterranean coastal waters

Xavier Ges,[1] Salvador Bará,[2,*] Manuel García-Gil,[1] Jaime Zamorano,[3] Salvador J. Ribas,[4,5] and Eduard Masana[5]

[1]*Departament de Projectes d'Enginyeria i la Construcció, Universitat Politècnica de Catalunya/BARCELONATECH, Barcelona, Spain.*

[2]*Facultade de Óptica e Optometría, Universidade de Santiago de Compostela, 15782 Santiago de Compostela, Galicia.*

[3]*Dept. de Astrofísica y CC. de la Atmósfera, Fac. de Ciencias Físicas, Universidad Complutense, Madrid, Spain.*

[4]*Parc Astronòmic Montsec, Camí del coll d'Ares s/n, 25691 Àger, Lleida, Spain.*

[5]*Departament de Física Quàntica i Astrofísica, Universitat de Barcelona, Barcelona, Spain.*

[*]*Corresponding author: salva.bara@usc.es*

**Abstract**

Light pollution is a worldwide phenomenon whose consequences for the natural environment and the human health are being intensively studied nowadays. Most published studies address issues related to light pollution inland. Coastal waters, however, are spaces of high environmental interest, due to their biodiversity richness and their economical significance. The elevated population density in coastal regions is accompanied by correspondingly large emissions of artificial light at night, whose role as an environmental stressor is increasingly being recognized. Characterizing the light pollution levels in coastal waters is a necessary step for protecting these areas. At the same time, the marine surface environment provides a stage free from obstacles for measuring the dependence of the skyglow on the distance to the light polluting sources, and validating (or rejecting) atmospheric light propagation models. In this work we present a proof-of-concept of a gimbal measurement system that can be used for zenithal skyglow measurements on board both small boats and large vessels under actual navigation conditions. We report the results obtained in the summer of 2016 along two measurement routes in the Mediterranean waters offshore Barcelona, travelling 9 and 31.7 km away from the coast. The atmospheric conditions in both routes were different from the ones assumed for the calculation of recently published models of the anthropogenic sky brightness. They were closer in the first route, whose results approach better the theoretical predictions. The results obtained in the second route, conducted under a clearer atmosphere, showed systematic differences that can be traced back to two expected phenomena, which are a consequence of the smaller aerosol content: the reduction of the anthropogenic sky glow at short distances from the sources, and the slower decay rate of brightness with distance, which gives rise to a relative excess of brightness at large distances from the coastline.

\*\*\*\*\*\*\*\*\*\*\*\*\*\*

## 1. Introduction

Light pollution is a widespread phenomenon whose consequences can be detected at places located hundreds of kilometers away from the light polluting sources [1-4]. A significant effort has been devoted in the last years to develop and validate quantitative models of the anthropogenic light propagation through the atmosphere [5-10]. Light pollution has been recognized as a relevant environmental stressor, and a large variety of ecological effects of artificial light at night have been reported in recent works [see, e.g. 11-17].

Although most research in this field has traditionally addressed the unwanted effects of artificial light inland, there is a growing interest in the study of light pollution in coastal regions, due to both the relevance of the marine areas for the preservation of biodiversity, and the opportunity they present for validating or otherwise contradicting physical models of atmospheric propagation with relatively simple source distributions.

Coastal regions are zones of interaction of sea and land processes, and home to a high number of species. Due to the high population density along the shoreline, this strategic region for biodversity is under strong pressure from anthropogenic polluting agents, including artificial light at night [18-20]. Although we are still far from having a complete picture of the complex effects induced by artificial light in the marine environment, several relevant impacts have been reported and are well documented. They include changes in the zooplankton diel vertical migration, bird collisions with ships, intensified visual predation and foraging in illuminated shorelines, sessile larvae settlement disruption, desynchronization of reproductive behaviour, nesting site displacement, and streetlight-induced disorientation in seaward or long-range migration of several species [18, 21]. In areas close to the light emitting sources the situation under overcast skies is aggravated, since the cloud cover acts as a powerful skyglow amplifier, due to the direct reflections of the light in the base of the clouds [4, 22-25]. Impacts to the ecosystem can be caused by very low light levels, even smaller than full-moon illuminance (that is, values below 0.26 lx [26]). Such low levels can be detected tens of kilometers away from big cities [27].

On the other hand, the marine environment provides a relatively simple stage for checking the performance of light pollution propagation models. If the sea region under study is relatively free from local light sources (offshore platforms, anchored vessels, heavy marine traffic or working fishing vessels), the zenithal sky brightness for constant atmospheric conditions is determined by the distribution of inland lights, and its dependence on the distance to the shoreline can be experimentally studied with relative ease. Furthermore, there are no screening factors that could complicate the analysis as those commonly found inland, as e.g. light blocking by obstacles like trees, buildings, or local topographic features.

In a recent study, Jechow et al [28] reported the use of fisheye DSLR cameras for characterizing the all-sky brightness about one mile offshore in the Gulf of Aqaba, near Eliat, Israel. Spatially-resolved measurements of the anthropogenic light propagation through the water column, up to 30 meters in depth, have been carried out in the same waters by Tamir et al [29]. Quantitative comparisons of atmospheric propagation models in island settings have been previously reported by Aubé and Kocifaj [30]. Observational studies of the spatial distribution of the zenithal night sky brightness inland have been carried out in recent years, among others, by Biggs et al [31], Pun and So [32], Espey and McCauley [33], Sánchez de Miguel [34], and Zamorano et al [35].

We describe in this work a design concept of a zenithal sky brightness measurement platform specifically devised for use on ships, under a wide range of navigation conditions, and present the results of two measurement runs from near the Barcelona shoreline, travelling up to 9 and 31.7 km into the Mediterranean, respectively. The measured brightness can be compared with the predictions of *The new world atlas of artificial night sky brightness* of Falchi et al [2], henceforth referred to as the NWA. A reasonably good overall agreement is found for comparable atmospheric conditions, as those prevailing in the first route. The second route was carried out under an air column with smaller aerosol content, and the qualitative behaviour of the artificial sky glow with distance follows the expected trends, that is, a reduction of the anthropogenic sky glow at short distances from the sources, and a slower decay rate of brightness with distance, giving rise to a relative excess of brightness at large distances from the coastline, in comparison with the NWA predictions.

## 2. Materials and Methods

### A. Light sensor

Zenithal night sky brightness measurements were made with a SQM-LU-DL light meter from Unihedron (Ontario, Canada). This low-cost device is based on a TSL237 high-sensitivity irradiance-to-frequency converter, with temperature correction (TAOS, USA). A concentrator optic restricts the device field of view to a region of the sky with approximately Gaussian transmittance profile with a full-width at half maximum of 20° [36]. A Hoya CM-50 filter limits the spectral band to 400-650 nm (effective passband of the whole setup at 50% of the maximum sensitivity) [36-37]. The brightness within a given photometric band is usually expressed in the negative logarithmic scale of *magnitudes per square arcsecond* (mag/arcsec$^2$), a non SI unit for the integrated spectral radiance, scaled and weighted by the spectral sensitivity of the detector [4]. Due to the particular spectral sensitivity of the SQM device, the conversion between the SQM magnitudes and the corresponding mag/arcsec$^2$ in standard photometric bands, like those defined by the Johnson-Cousins *B*, *R* or *V* filters [38], or the CIE photopic visual spectral efficiency function V($\lambda$) [39], turns out to be spectrum-dependent, and in the absence of *a priori* information it can be done only in an approximate way. To avoid confusion, we will henceforth denote by mag$_{SQM}$/arcsec$^2$ the units of the measurements directly provided by the SQM sensor, and by mag$_V$/arcsec$^2$ the units of the measurements in the Johnson's *V* photometric band. A widely-used approximate formula for converting any given value $m_V$ in mag$_V$/arcsec$^2$ to luminance *L* in SI units cd·m$^{-2}$ is [4-5]

$$L[cd \cdot m^{-2}] = 10.8 \times 10^4 \times 10^{(-0.4 m_V)} . \qquad (1)$$

Note that dividing this luminance by the overall luminous efficacy factor 683 lm/W we get the corresponding radiance (in Wm$^{-2}$sr$^{-1}$) present at the output of the V-filter [4, 39].

## B. Gimbal mount

In order to keep the device pointing to the zenith under actual navigation conditions it is necessary to compensate for the pitch and roll rotations of the ship around its principal axes. The magnitude of these rotations is dependent on the ship geometry and dynamics, and on the local weather and wave slope conditions. In order to preserve the vertical alignment of the sensor, a dedicated two-axis gimbal (Cardan) mount was specifically designed for this project (Fig. 1). The enclosure was designed to withstand maximum ship transverse inclinations (roll) of ±50°, and longitudinal ones (pitch) of ±25°. The SQM housing is a freely oscillating system, and its ability to preserve the vertical orientation is optimized when the sea-wave frequency, as seen from the ship reference frame, is far from the resonant frequency of the mechanical set up (a few Hz). The vertical alignment was designed to be kept within ±5° for accelerations smaller than 0.85 m s$^{-2}$. The mount dimensions are 454 mm x 404 mm x 330 mm, the overall weight is 5.6 kg, and it provides IP67 protection against dust and water [40] with working temperature range −5 to 45 °C.

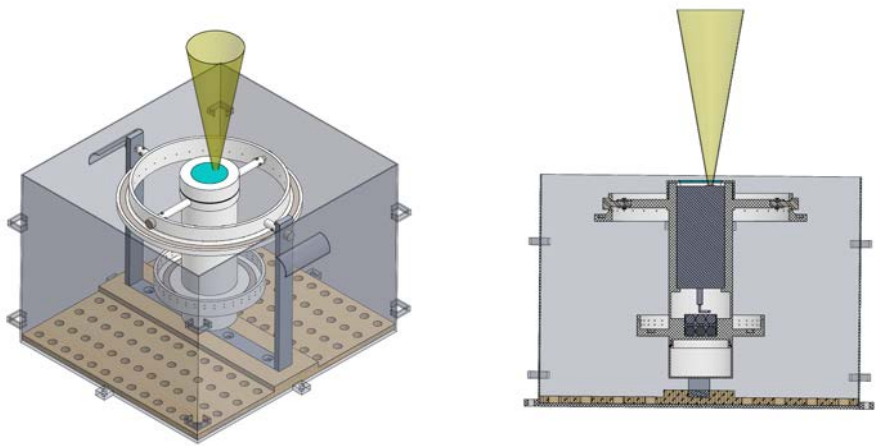

**Figure 1:** Gimbal mount design holding the SQM housing.

The prototype actually built and used in the measurement routes had some minor differences with the one originally designed. The outer casing was of slightly smaller dimensions, and the effective working ranges were ±40° (roll) and ±30° (pitch). The SQM housing cylinder was supplied by the manufacturer, and properly compensated to keep balanced the center of mass.

Although this gimbal mount can withstand significant vessel inclinations, the night sky brightness measurements were taken in optimum sea conditions: without swell, navigation at proper and constant speed for the characteristics of the ship and weather conditions, only propelled by a four-stroke gasoline outboard engine. The objectives were to guarantee the vertical pointing of the SQM, and the soft testing of the device.

*C. Sea routes*

Two measurement routes were followed in the nights of 2 to 3 July, and 7 to 8 August 2016, respectively, the first one starting from the harbor of Barcelona, north entrance, travelling perpendicularly to the coast for a distance of 9 km, and the second one starting from the nearby municipality of Badalona, travelling 31.7 km (Figure 2). The measuring platform was installed in a Mini Transat 6.50 Pogo 1 sailing vessel of 6.50 m length, 2.99 m width and 1.80 m draught, with an 11-m mast and an auxiliary 5 HP engine, providing a maximum motor speed of 9 km/h (5 knots). Zenithal sky brightness measurements were taken at a rate of one every fifteen seconds, which, at maximum motor speed, provides a spatial sampling resolution of 37.5 m. The local coordinates were acquired by a GPS at a rate of 1 per second, and stored in the computer. The distance to the starting point was determined using the well-known haversine formula [41]. The main parameters of the routes, and the sky and sea conditions of each journey, are summarized in Table 1 (see section 3).

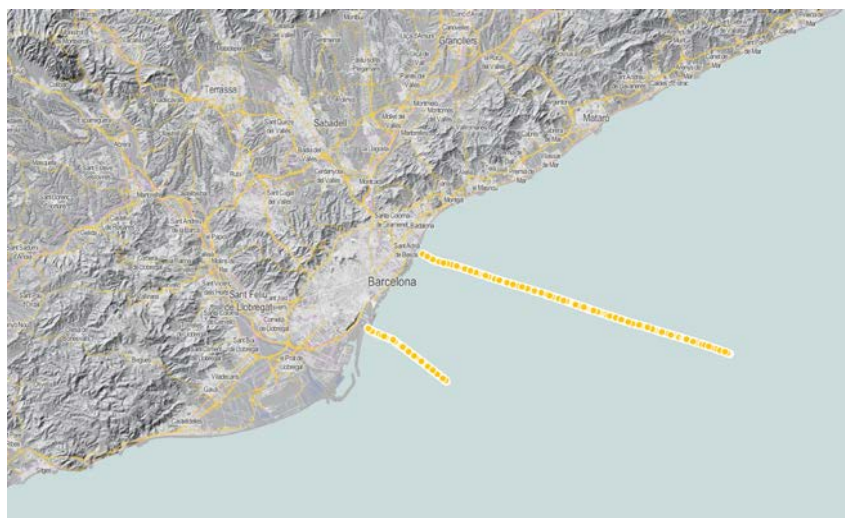

**Figure 2:** Measurement routes of the nights of 2 to 3 July 2016 (lower track) and 7 to 8 August 2016 (upper track). Cartography CC-BY Institut Cartogràfic de Catalunya / OSM.

*D. Predicted sky brightness and data processing*

The floating point dataset of the NWA [42] provides detailed predictions of the artificial component of the zenithal sky brighness for all regions of the world, with 30 arcsec ground spatial resolution. At the latitude of Barcelona this resolution corresponds to a spatial pixel of size 926 m (along the local latitude axis) x 695 m (along the local longitude one). The NWA data are given in SI units of millicandelas per square meter (mcd/m$^2$). The predicted artificial sky brightness for each point of our routes was obtained by extracting the values of the NWA geotiff file at the corresponding coordinates.

The SQM measurements recorded in our expeditions, however, include both the artificial and the natural components of the zenithal night sky brightness. In order to make meaningful comparisons with the predictions of the NWA, the natural component has to be subtracted from the measurements. The natural brightness of the sky in cloudless nights depends on a large set of factors, including the phase and position of the Moon, the Milky Way, other conspicuous celestial bodies (bright stars and planets), the zodiacal light, and the natural airglow. The overall contribution of these sources can be accounted for by different methods. One of them is to use an accurate model of the natural sky brightness, like the one developed by Duriscoe [43]. Another possibility, albeit only coarsely approximate, is to subtract from the measured values a standard constant offset, based e.g. on the expected brightness of the natural moonless sky away from the Milky Way, zodiacal light and bright pointlike sources, reported to be of order 0.174 mcd/m$^2$, equivalent to about 22.0 mag$_V$/arcsec$^2$ [2], or some sensible correction of this figure depending on the detailed sky conditions. A third option, the one we used in this work, is to estimate the natural sky brightness present in our measurements by using the zenithal brightness data acquired with the same kind of sensor in a pristine dark place, located not too far from the boat position, during the same or nearby nights, and under comparable atmospheric conditions.

To that end we resorted to the measurements recorded by the SQM-LE sensor run by the Servei Català per la Prevenció de la Contaminació Acústica i Lumínica de la Generalitat de Catalunya and Parc Astronòmic Montsec, in the framework of a pilot plan of a Catalan Network, located at the Montsec Astronomical Observatory (*Observatori Astronómic del Montsec*, OAdM, 42.0517° N, 0.7296° E), a dark site in a mountain range about 60 km North of Lleida, Catalonia. In order to apply these data to subtract the natural sky contribution from the ship measurements, we took into account the difference of longitudes between the OAdM and the ship position along the sea routes. These longitude differences translate into relative delays in the time at which any celestial feature located at the zenith of the ship reaches the zenithal region in the dark place. The dark sky data gathered at the OAdM were compared with the readings of the SQM-LE sensor located at the Centre d'Observació de l'Univers (COU) del Parc Astronòmic Montsec (42.0247° N, 0.7368° E), as well as with the ones recorded by the SQM-LE sensor of the Spanish Network for Light Pollution Studies [44] at the neighbouring site of Cal Maciarol (42.0179° N, 0.7441° E). All datasets provided comparable levels of correction for the natural dark-sky signal (the inter-device absolute differences were in all cases smaller than 0.1 mag$_{SQM}$/arcsec$^2$). The Cal Maciarol data were not available for the very same night of the first route, and data from neighboring nights were used, correcting the time of the zenital culmination of a given patch of the sky for the sidereal rotation of the Earth (3 min 56 sec of advance per elapsed day). The readings, originally given in mag$_{SQM}$/arcsec$^2$, were transformed into equivalent luminances using Eq. (1). In such way we obtained the luminances $L_{raw}(t)$ and $L_{nat}(t)$, corresponding to the ship and the OAdM measurements, respectively. Note that this introduces some error if the results are interpreted as true luminances in cd/m$^2$, because the SQM and V photometric bands are not exactly the same [49-50]. The natural sky luminance was then subtracted from the total one recorded from the ship at each point of the route, allowing us to estimate the value of the artificial component as $L_{raw}[\mathbf{r}(t)] - L_{nat}(t')$, where **r**(t) is the position of the ship at time *t*, and *t'* is the time at which the objects located at *t* in the zenith of the ship reached the zenith at the dark site, that is, the time *t* corrected for the difference of longitudes (and, possibly, of measurement dates) between this site and the ship.

The NWA calculations were based on the upward artificial light emissions measured by the VIIRS-DNB sensor on board the Suomi-NPP polar-orbiting satellite, and were calibrated using SQM readings recorded by observers distributed throughout the globe, including a relevant data set from the Montsec area [2]. The VIIRS-DNB images are taken each night at about 01:30 local mean solar time, in our case within the UTC zone. They correspond, then, to the anthropogenic emissions at this particular moment of the night. As a matter of fact, the global urban light emissions are not constant but tend to decrease at a variable rate throughout the night (roughly of order 4.5% per hour [2]). To correct our artificial skyglow measurements at the sea for the variable amount of urban emissions, we used the data recorded by the SQM-LE sensor installed on the rooftop of the Physics Faculty of Universitat de Barcelona (41.3847° N, 2.118° E). The measurements done by this sensor, located within a brightly lit metropolitan area at the starting point of the routes, allow us to estimate with sufficient degree of accuracy the time dependence of the artificial light emissions in the area that most contribute to the measurements made from the ship. The natural sky brightness in Barcelona at time $t$ was estimated from the OAdM data $L_{nat}(t'')$ at corrected $t''$ times, following the same procedure described above, and subtracted from the raw values. As expected, the natural component of the night sky at the Barcelona metropolitan area is fairly negligible in comparison with the artificial one. The resulting data can be directly used to assess the relative changes over time of the artificial component of the sky brightness at the ship position. We transformed the urban data, originally given in $mag_{SQM}/arcsec^2$, into equivalent luminances $L_{urban}(t)$ using Eq. (1). The same remark made above about the differences between the SQM and V photometric bands applies, as well as the caution needed to interpret the results in terms of SI $cd/m^2$. The measurement times of this sensor were not adjusted for the differences of longitudes between this urban site and the ship locations, since the urban light emissions are mainly dependent on the standard time (CEST=UTC+2) and not on the times at which some celestial regions may reach the local zenith.

The artificial night sky brightness at each point **r** of the measurement route, $L_{artif}[\mathbf{r}(t)]$, normalized to the urban emissions at $t_0$=01:30 UTC, was then estimated as:

$$L_{artif}[\mathbf{r}(t)] = [L_{raw}[\mathbf{r}(t)] - L_{nat}(t')]\left[\frac{L_{urban}(t_0=01:30) - L_{nat}(t''_0)}{L_{urban}(t) - L_{nat}(t'')}\right], \qquad (2)$$

where $t''_0$ is the time $t_0$ corrected for the difference of longitudes between Barcelona and the OAdM.

## 3. Experimental results

Table 1 summarizes the main parameters of the two measurement routes carried out in the summer of 2016, whose geographic paths are displayed in Figure 2 above. Both routes presented favourable atmospheric and sea conditions, with the Moon below the horizon during the effective measurement period.

The readings of the SQM on board the ship were corrected for the transmittance of the glass window of the housing by subtracting 0.11 arcsec$^2$ from the raw values provided by the sensor. The resulting SQM measurements (henceforth SQMshp) were calibrated, i.e. converted into equivalent readings of the reference SQM detector of the Catalonian light pollution measurement network (henceforth SQMxcl) using an appropriate linear regression model [45]. To that end, the SQM used aboard the ship was installed at the COU, Parc Astronòmic Montsec, during the summer of 2017, taking continuous measurements in parallel with several calibrated detectors of the Catalan Network. The effect of the ship navigation lights was assessed in a place far from the coast, at 41.344° N and 2.552° E, under dark skies comparable to the darkest ones in the routes. The recorded sky brightness with the navigation lights and screens switched on was 21.02 mag$_{SQM}$/arcsec$^2$, darkening to 21.12 mag$_{SQM}$/arcsec$^2$ with the fixtures switched off. According to Eq. 1 the ship lights contributed with 0.0426 mcd/m$^2$ to the SQM readings, which were subtracted from the measured raw luminances during the data processing workflow. The OAdM and the Universitat de Barcelona SQM readings were also calibrated according to their corresponding XCL protocols. The time dependence of the SQM brightness at these three sites during the two measurement routes is displayed in Fig. 3. Both the SQMshp and the SQMxcl data of the ship sensor are shown. Note that the data of the Barcelona urban site were unavailable for the night of route 1, and data taken one day later were used instead. All sensors provided reliable data in real time for route 2. A detailed view of the time dependence of the natural zenithal sky brighness measured in the OAdM is displayed in Fig. 4.

**Table 1: Measurement routes**

| Parameter | Route 1 | Route 2 |
|---|---|---|
| Date | Night 2 to 3 July 2016 | Night 7 to 8 August 2016 |
| Time period | 00:29 - 01:44 CEST | 23:37 - 03:28 CEST |
| Sun altitude | −22° to −25° | −24° to −29° |
| Moon altitude | −31° to −29° | −2° to −42° |
| Moon illuminated surface | 3.4% | 23.8% |
| Cloud conditions | clear | clear |
| Maximum wave height | 0.7 m | 0.5 m |
| Maximum wind speed | 10 knots | 7 knots |
| Starting point | 41.355°N / 2.180°E | 41.425°N / 2.236°E |
| Course | 125° | 115° |
| No. of measurements | 302 | 924 |
| Measurement interval | 15 s | 15 s |
| Distance travelled | 9 km | 31.7 km |
| Aerosol optical depth $\tau_a$ (*) | 0.17 | 0.09 |

(*) Data from the AERONET station in Barcelona [47], corresponding to the averages of the days 2-3 July and 7-8 August, respectively.

The boat measurements were corrected for the natural background and the changing urban emissions using Eq. (2). Figure 5 shows the results of these corrections.

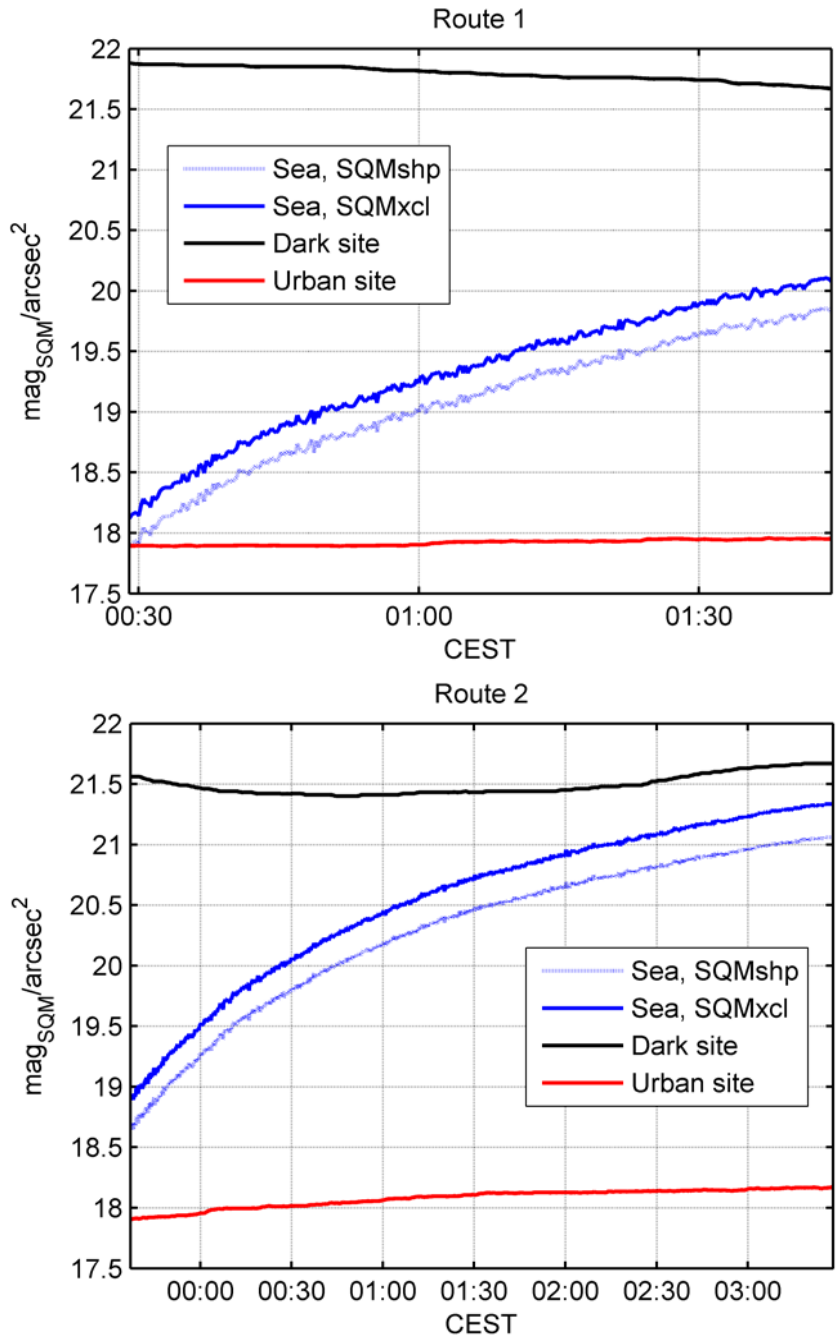

**Figure 3**: Time dependence of the measured zenithal sky brightness ($mag_{SQM}/arcsec^2$) at the boat location (SQMshp and SQMxcl), OAdM dark site, and urban center of Barcelona.

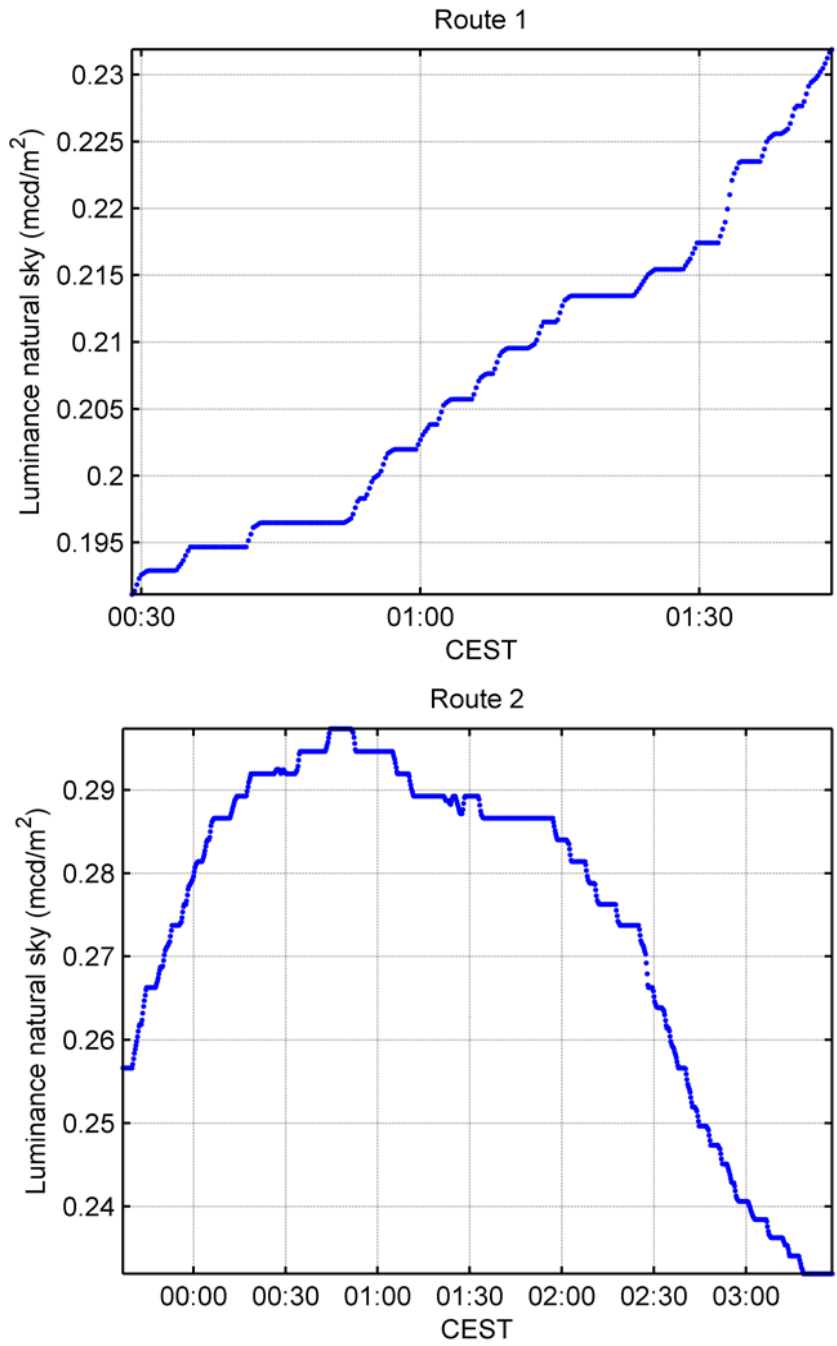

**Figure 4**: Time dependence of the natural zenithal sky brightness, $L_{nat}(t)$, in mcd/m², see Eq.(1), evaluated at the ship position, using the OAdM dark site data at times corrected for the difference of longitudes between the OAdM and the ship.

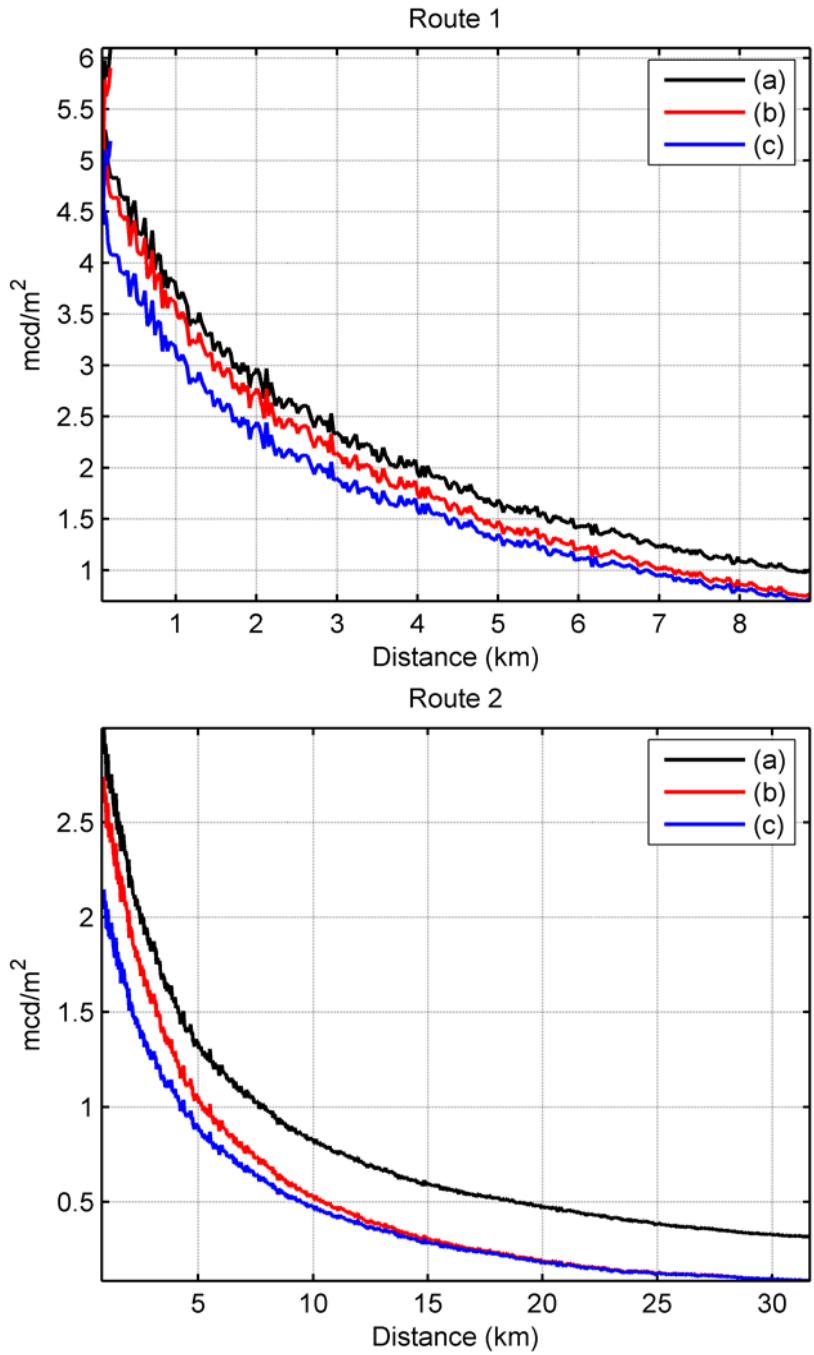

**Figure 5**: Zenithal sky brightness measurements (SQMxcl) at the ship location, in mcd/m$^2$, see Eq.(1), versus distance (km) to the starting point of each maritime route. (a) total brightness, $L_{raw}(t)$; (b) artificial component, $L_{raw}[\mathbf{r}(t)] - L_{nat}(t')$; (c) artificial component normalized to constant urban emissions at 01:30 UTC, $L_{artif}[\mathbf{r}(t)]$.

The dependence of the artificial sky brightness $L_{artif}(\mathbf{r})$ on the distance to the coast, as deduced from our measurements and the NWA predictions, is shown in Fig. 6 and 7, in luminance (mcd/m$^2$) and mag/arcsec$^2$ units, respectively. Fig. 8 and 9 display the scatter plots of both measurement datasets. The NWA plots show a staircase structure, due to the fact that the NWA spatial resolution is coarser than the one of the sea measurements: the combination of the ship speed and the SQM measurement rate resulted in several measurements being taken within each NWA pixel.

Table 2 shows the residual differences between the measured and the predicted values of the artificial sky brightness. The rms difference in mag/arcsec$^2$ of the raw measurements is below 0.1 for route 1, and lies within 0.3 mag/arcsec$^2$ for route 2. Once calibrated to the XCL reference, these differences increase to 0.34 and 0.68 mag/arcsec$^2$, respectively.

**Table 2:** Residual differences between the boat measurements and the NWA estimations of the artificial sky brightness, in luminance (mcd/m$^2$) and mag/arcsec$^2$ units.

| Differences | Route 1 | | Route 2 | |
|---|---|---|---|---|
| **SQMshp** | mcd/m$^2$ | mag/arcsec$^2$ | mcd/m$^2$ | mag/arcsec$^2$ |
| mean | −0.101 | 0.058 | −0.259 | 0.165 |
| stdev | 0.218 | 0.065 | 0.431 | 0.252 |
| rms | 0.240 | 0.087 | 0.502 | 0.301 |
| **SQMxcl** | mcd/m$^2$ | mag/arcsec$^2$ | mcd/m$^2$ | mag/arcsec$^2$ |
| mean | −0.608 | 0.332 | −0.445 | 0.669 |
| stdev | 0.296 | 0.074 | 0.538 | 0.097 |
| rms | 0.676 | 0.340 | 0.698 | 0.676 |

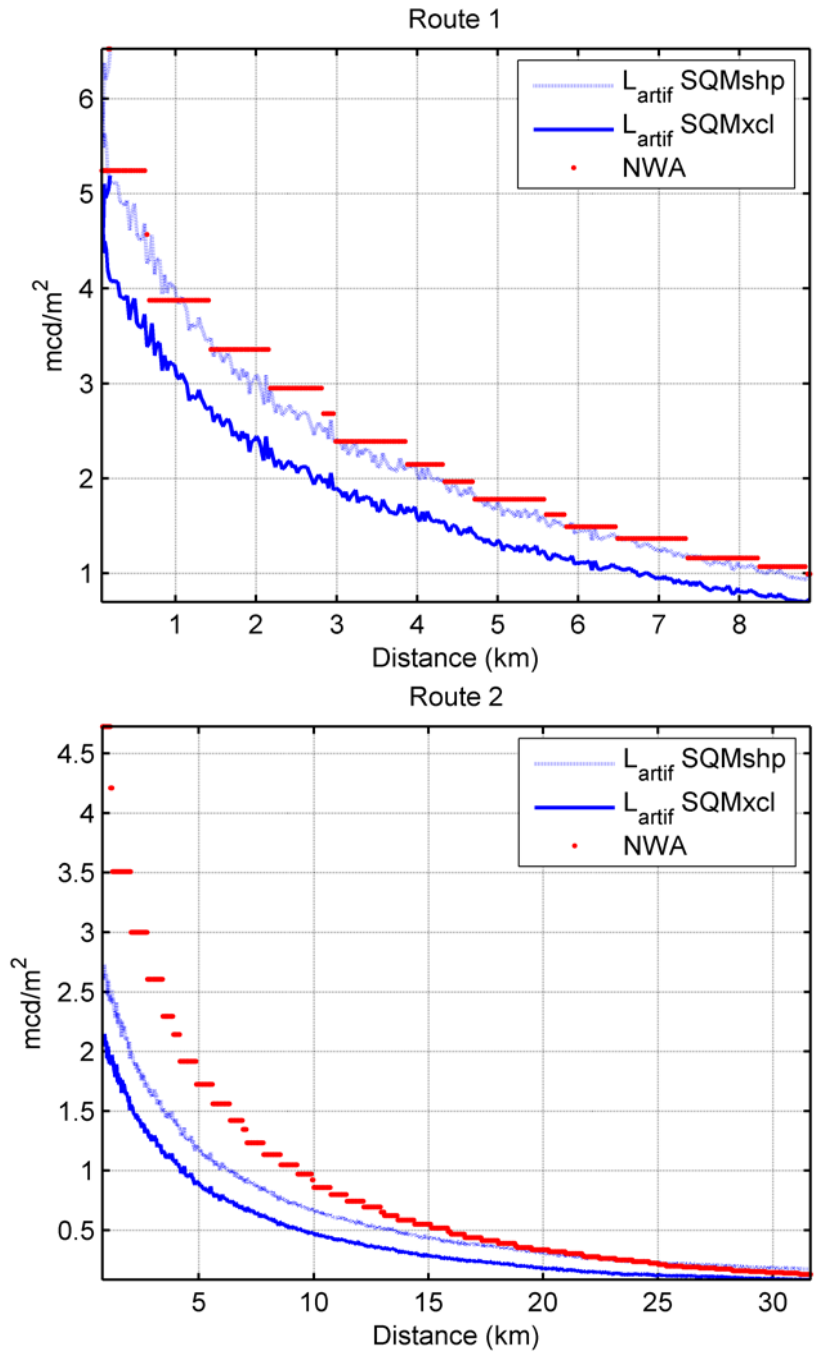

**Figure 6**: Measured and predicted artificial zenithal night sky brightness (mcd/m$^2$) vs distance to the starting point of each route. The figure shows the sea measurements corrected for constant urban emissions, $L_{artif}(\mathbf{r})$, obtained from SQMshp and SQMxcl, and the NWA predictions.

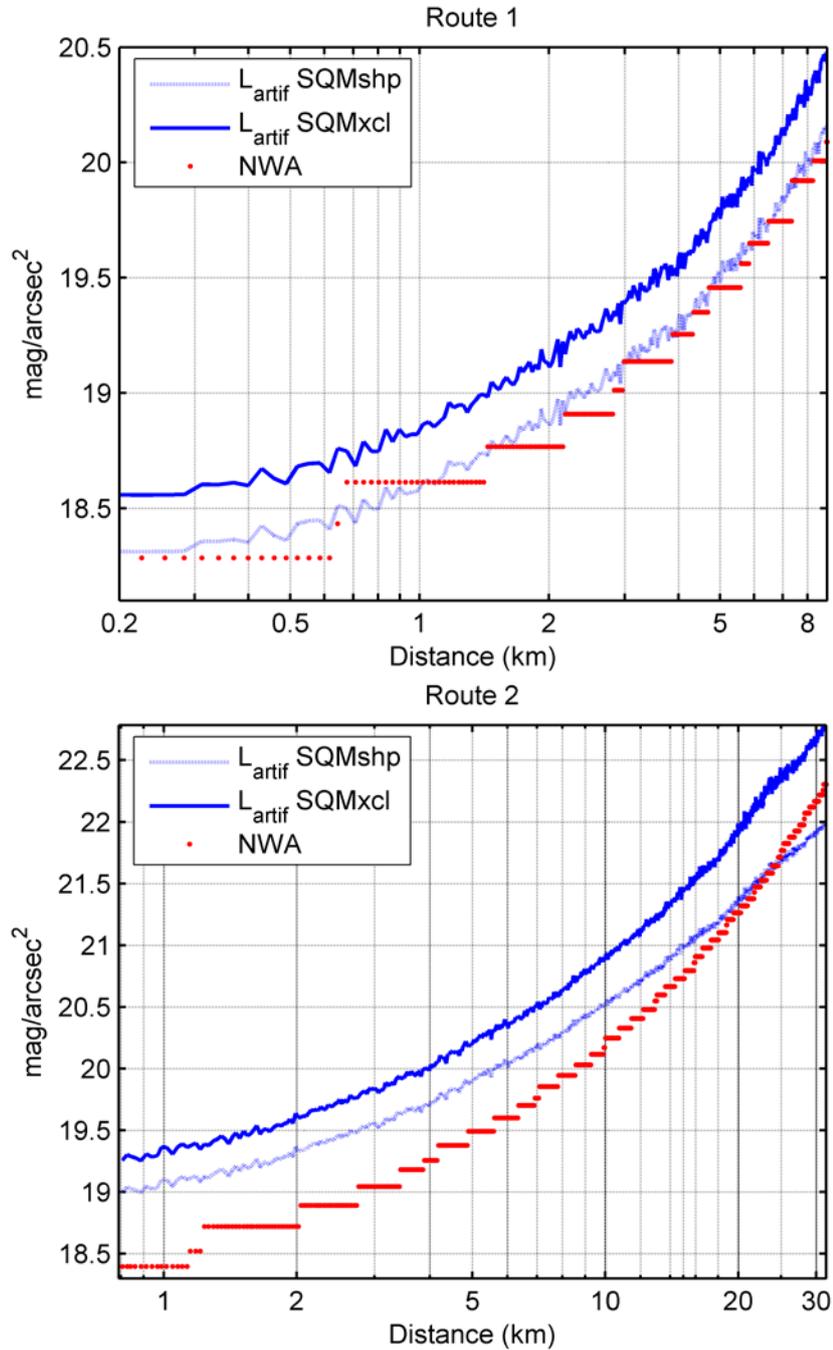

**Figure 7**: Measured and predicted artificial zenithal night sky brightness (in mag/arcsec$^2$) vs the distance to the starting point of each route, in logartihmic scale. The figure shows the sea measurements corrected for constant urban emissions, $L_{artif}(\mathbf{r})$, obtained from SQMshp and SQMxcl (mag$_{SQM}$/arcsec$^2$), and the NWA predictions (mag$_V$/arcsec$^2$). If the brightness (in cd/m$^2$ or Wm$^{-2}$sr$^{-1}$) varied with distance (km) according to a power law, the resulting plots would be straight lines.

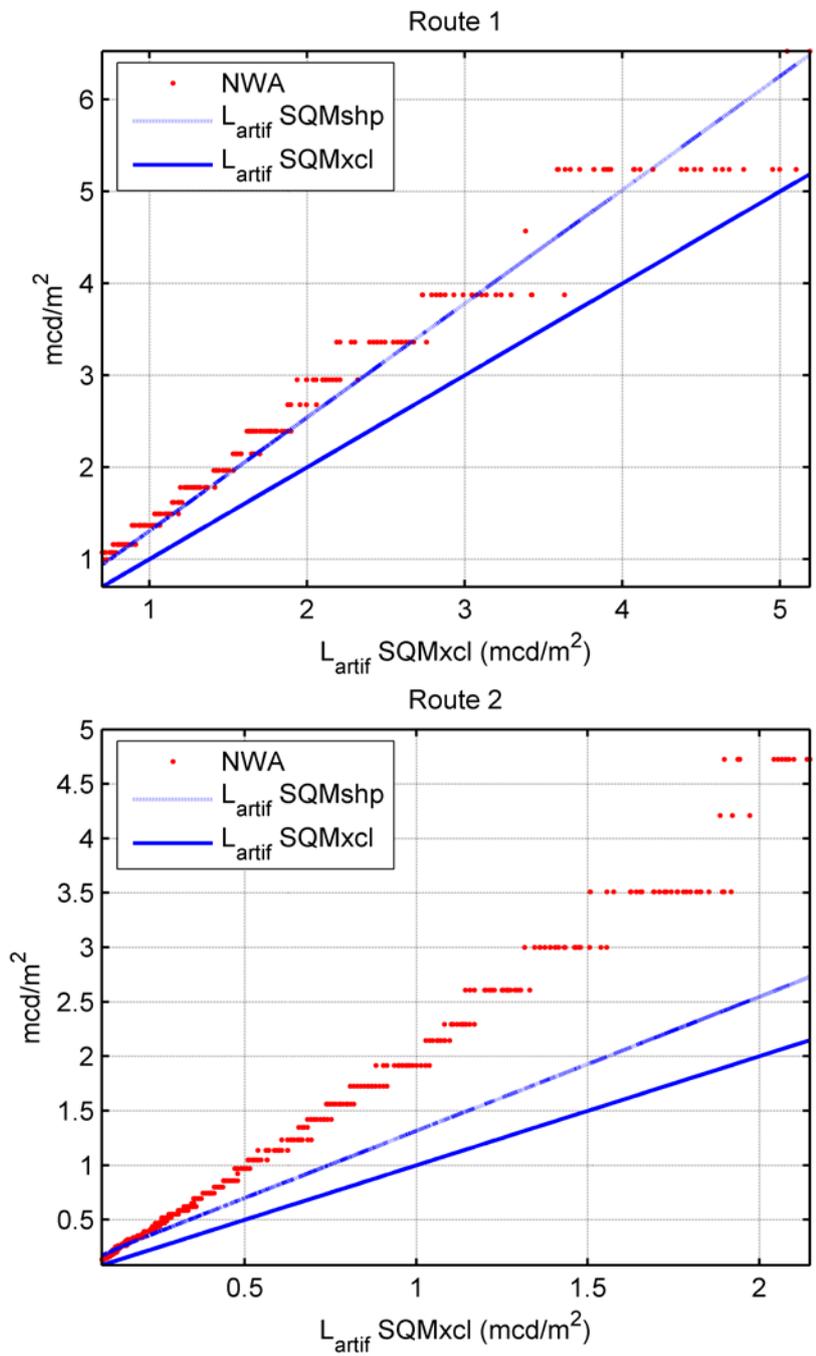

**Figure 8**: NWA vs $L_{artif}(\mathbf{r})$ luminance (mcd/m²) computed from SQMxcl, for constant urban emissions. The solid line (L$_{artif}$ SQMxcl) is the unit slope straight line passing through the origin.

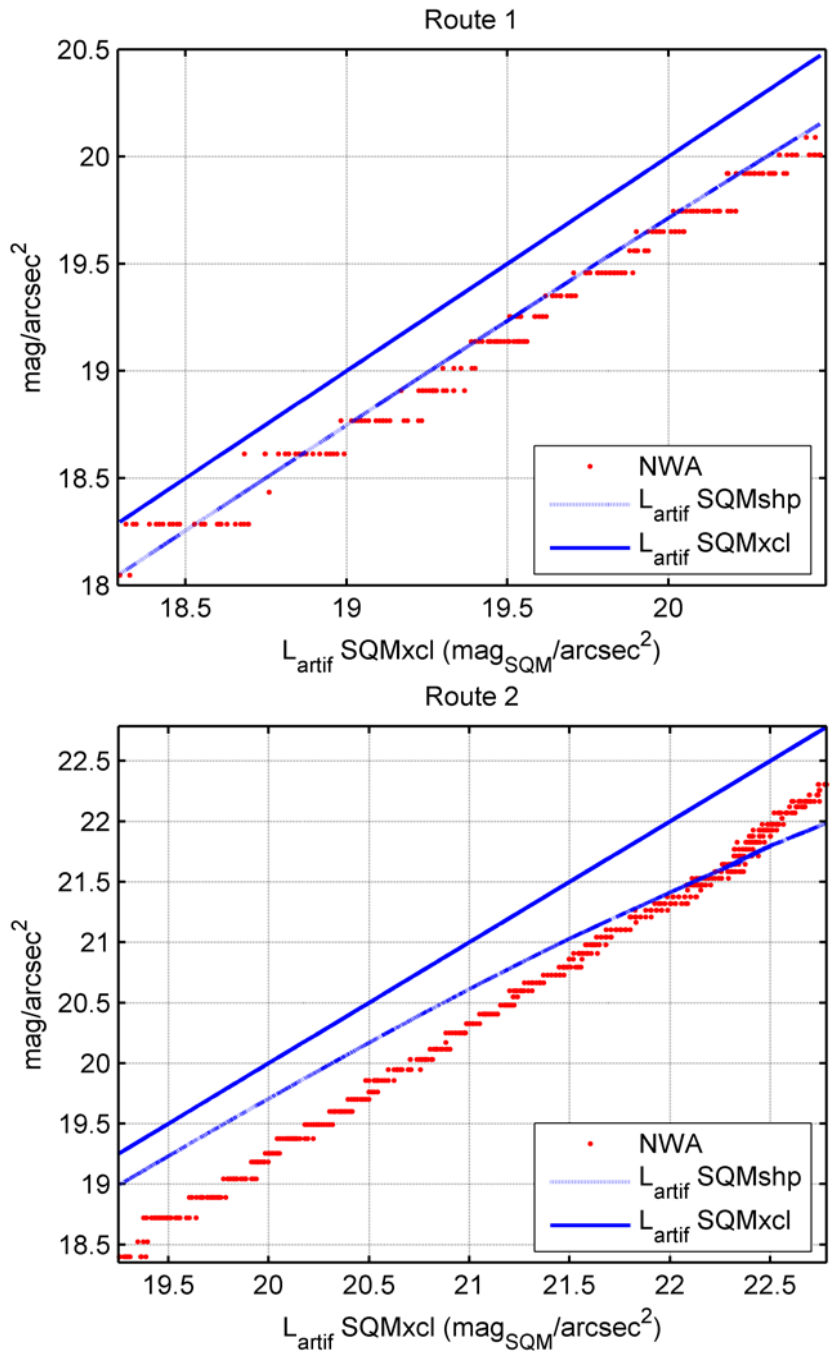

**Figure 9:** NWA vs $L_{artif}(\mathbf{r})$ luminance (mag/arcsec$^2$) computed from SQMxcl, for constant urban emissions. The solid line (L$_{artif}$ SQMxcl) is the unit slope straight line passing through the origin.

## 4. Discussion

The SQMshp measurements obtained in route 1 (Figs. 6-7, top) show a remarkable degree of coincidence with the predictions of the NWA [2,42]. These predictions were computed using the method of Cinzano et al [46], assuming a standard clear US62 atmosphere, and an aerosol clarity parameter $K=1$ [5]. This value of $K$ corresponds to an aerosol optical depth $\tau_a=0.19$, a vertical extinction of 0.33 $mag_V$ at sea level, horizontal visibility of 26 km, and a total atmospheric optical depth (including molecular scattering and absorption, besides the aerosol ones) of $\tau=0.31$ [2]. During our first measurement route, the aerosol optical depth measured at the Barcelona AERONET station [47] was 0.23 and 0.11 for the days of 2 and 3 July, respectively, averaging to $\tau_a=0.17$. This corresponds to a clarity parameter $K=0.86$, with vertical extinction at sea level 0.31 $mag_V$, horizontal visibility 30 km, and a total atmospheric optical depth $\tau=0.28$. These atmospheric conditions were reasonably close to the ones assumed in the NWA calculations.

The SQMshp measurements obtained in route 2, in turn, show systematic differences with respect to the NWA predictions (Figs 6-7, bottom). The atmosphere, however, had a substantially lower aerosol content that night, averaging to $\tau_a=0.09$ according to the AERONET records at Barcelona (0.11 and 0.07 for the days of 7 and 8 August, respectively). This corresponds to a clarity index $K=0.45$, vertical extinction 0.22 $mag_V$, horizontal visibility 53 km and total atmospheric optical depth $\tau=0.20$. The overall effect of a reduced aerosol content (which corresponds to smaller values of $K$) on the artificial skyglow is described in Figure 3 of Ref. [5]. If the aerosol content is smaller, the artificial skyglow decreases close to the sources, because the scattering efficiency is correspondingly smaller, but tends to increase at longer distances, since the higher transmittance of the atmosphere gives rise to higher incident radiances that finally overcome the smaller scattering rate. This general prediction of the numerical models is qualitatively supported by our experimental data. Note however that no further quantitative statements can be made in the absence of NWA numerical predictions corresponding to these particular atmospheric conditions.

The SQMxcl calibrated data significantly depart from the NWA predictions in both routes. The readings of the SQM sensor on board of the boat were in average 0.25 mag$_{SQM}$/arcsec$^2$ brighter than the ones the Catalonian network reference SQM sensor would record. This systematic difference is significantly larger than the expected 0.1 mag$_{SQM}$/arcsec$^2$ inter-device reproducibility of the SQM sensors according to manufacturer's specifications. When this bias is taken into account, the measured brightness is always smaller than the predicted one. The overall trends observed in the raw data are still preserved, however, with the first route showing a closer agreement between the measurements and the NWA expected values. The difference between the SQMshp and the SQMxcl values is intriguing. The Catalonian Network (XCL) reference has been established from the average readings of several SQM detectors working in parallel under well controlled conditions [45]. This average detector has been shown to be consistent with the readings of wider sets of SQM devices in extensive intercomparison campaigns [48]. The SQM detector used on the ship, on the other hand, is not an isolated outlier. Several SQM sensors of the 27xx series provide compatible interdevice readings, showing similar differences (of order 0.2 mag$_{SQM}$/arcsec$^2$) with respect to the XCL standard. This difference in magnitudes amounts to a multiplicative factor in the associated luminances. A related issue is what calibration is the most adequate to compare SQM readings against the NWA predictions. Note that the conversion from SQM magnitudes to luminances given in Eq.(1) is only approximate, because any accurate conversion depends on the detailed spectral composition of the light scattered downwards from the zenith [49]. Besides, the SQM band (the one used in the measurements and the overall calibration of the NWA), the Johnson-Cousins V (used in the computation of the NWA), the VIIRS-DNB band (providing the artificial light emissions data for the NWA predictions) and the visual CIE V($\lambda$) (that establishes the luminance scale) are not photometrically equivalent [34,49-50].

The relative blue insensitivity of the VIIRS-DNB radiometer is a relevant issue that shall be kept in mind when performing these comparisons. The SQM detector is fairly sensitive to the blue region of the optical spectrum, and this could lead the NWA to underestimate the SQM measurements in areas where there is a noticeable amount of high CCT LED source emissions. Besides, the differences between the VIIRS-DNB and the SQM spectral bands could lead to different brightness estimations even if the sources' emissions were fully comprised within both passbands. There are, however, two factors that, in our opinion, may partially attenuate this potentially disturbing bias. On the one hand the Barcelona streetlight census is still strongly based on High Pressure Sodium vapor sources (HPS): according to recently published data provided by the municipality, as of 2014 a 75% of Barcelona streetlights were HPS, versus 5% LED (the remaining ones being metal halides, residual Hg vapor, and others), while for the year 2020 the forecast is that HPS will still represent a noticeable share of 65%, with LED increasing up to the 20% of the total. It can be guessed that LED streetlights were less than the 10% of the total number of sources at the time our maritime routes were made in the summer of 2016. Since LED luminaires generally provide a better control of the angular distribution of the emitted light, it can be expected that their contribution to the artificial sky brightness was, at the time of our measurements, well below the 10% level. On the other hand, the NWA calibration was made using -among other inputs- a huge database of SQM measurements made by professional and citizen scientists throughout the world, a significant fraction of which were taken precisely in Catalonia. Since the calibration was based on getting the best fit of the overall NWA predictions to the SQM measured values, it is expected that the systematic bias between both datasets is reasonably small (see Fig. 18 of [2], error histogram for Catalonia).

The calibration of the NWA, which also resorted to photometric data from the US National Park Service acquired in dark locations, seems to suggest that the brightness recorded by the SQM (in linear units) for very dark sites is about 1.15 times the corresponding SI luminance [2]. If we apply this factor to our measurements, the coincidence between the measured and predicted brighnesses worsens for route 1 because it translates into an additional 0.15 $mag_V/arcsec^2$ offset, without providing any significant improvement for route 2. This offset does not modify the slope of the skyglow versus distance relationship of route 1 in $mag/arcsec^2$ units, which still coincides with the NWA predictions (Fig. 7, top). Note also that, according to manufacturer's specifications, the SQM inter-detector reproducibility is of order ±0.1 $mag_{SQM}/arcsec^2$, introducing an additional degree of uncertainty beyond which absolute radiometric comparisons may be difficult.

More research is needed, including instrument development for multispectral radiance sensing from low Earth orbit satellites, in order to elaborate quantitative predictions and collect field measurements in fully compatible spectral bands.

Determining the natural night sky brightness at the sea, a few tens of km away from a highly illuminated coast is a difficult challenge. In our case, the ideal and direct method of making sky brightness measurements before and after a full black-out of the metropolitan area of Barcelona was clearly unfeasible. The practicable approaches were either estimating the natural sky brightness from theoretical models or performing measurements in dark, nearly pristine locations, not too far away from the ship position. Either method has limitations that can compromise the estimation of the artificial component of the night sky brightness at long distances offshore, in the range of asymptotically small artificial brightness values (natural brightness corrections are of minor importance at short distances from the metropolitan areas, where the sky brightness is dominated, for all practical purposes, by the artificial emissions). The main sources of natural sky brightness are the celestial objects passing above the observer and the natural airglow. Both of them can be deemed approximately constant for places located about 150 km away along the same latitude, as it was the case of the ship and the Montsec observatory, once the appropriate correction for the different times at which any given patch of the sky reaches the zenith at each location (due to the difference of geographical longitudes) is performed. The natural sky brightness, however, does not depend only on its sources, but also on the transmittance properties of the atmosphere. Different aerosol concentration profiles could lead to additional biases in the estimation of the natural brightness at long distances from the coast. The bias due to a potentially different aerosol concentration profile at the Montsec site is difficult to estimate, since there are no available data from the Montsec AERONET station for the dates of our routes. All things considered, we used the dark-sky site measurement approach to estimate the natural brightness at the ship. The calculated artificial brighness values at the points of the second route located farthest away from the coast shall in consequence be taken with some caution.

Irrespective of overall scaling factors, the dependence of the artificial zenithal night sky brightness $L_{artif}(\mathbf{r})$, obtained either from SQMshp or SQMxcl readings, on the distance to the coastline recorded in our expeditions does not seem to be well described by a single power law relationship across the whole distance range. The same can be said of the NWA predictions for this particular region of the Mediterranean. If the brightness $L_{artif}(\mathbf{r})$ (in cd m$^{-2}$ or W m$^{-2}$ sr$^{-1}$) varied with distance *d* according to a power-law, the plots in Fig. 7 would be straight lines. Similar conclusions can be obtained regarding a tentative exponential relationship: the dependence of the logarithm of the night sky brightness (in cd m$^{-2}$ or W m$^{-2}$ sr$^{-1}$) with the distance *d*, not shown here, deviates significantly from a straight line if the distance range is sufficiently large.

The zenithal sky brightness sensors used at the sea have to compensate for the variable inclinations of the ship due to the prevailing atmospheric and swell conditions. This problem was addressed in our case by using a passive mechanical gimbal mount that allows significant attenuation of the ship movements. If the SQM versus time plot (blue line in Fig. 3) is fitted by a low degree polynomial and this overal trend is subtracted from the data, the residual random fluctuations amount to just 0.01-0.02 mag$_{SQM}$/arcsec$^2$, showing a remarkable stability of the measurements. Our proof-of-concept Cardan design was driven by the requirements of simplicity and low cost. Note, however, that several models of electronic Cardan mounts for professional and amateur use are available on the market at reasonable prices, and could represent an interesting and practical improvement over our purely passive system. A well-known application is their use for digital sports cameras.

## 5. Conclusions

We have developed a low-cost device which is, to the best of our knowledge, the first operational system for acquiring zenithal night sky brightness measurements in the sea under actual navigation conditions. The system is based on a SQM detector mounted on a passive mechanical gimbal, and its performance has been satisfactorily tested in two routes starting from the Barcelona metropolitan area, travelling 9 and 31.7 km into the Mediterranean along paths perpendicular to the coastline. The SQMshp measured brightness in the SQM spectral band was fairly coincident with the V-band estimates of the NWA in the first route, whose atmospheric conditions were relatively close to the ones used for the calculations of the NWA. The second route took place under an atmosphere with noticeably smaller aerosol content, and the dependence of the artificial zenithal sky glow with the distance to the coast followed the expected trend. After the SQMshp measurements were calibrated to the reference SQM of the Catalonian light pollution measurement network, the differences between the observed and the NWA predicted brightness increased by about 0.25 $mag_{SQM}/arcsec^2$, although the overall trends remained invariant.

The measurement of the night sky brightness across extended regions of the sea may provide the necessary data for validating the current numerical predictions of artificial skyglow based on satellite observations and atmospheric scattering models. Ample regions of the sea surface are an environment free from artificial sources and irregular obstacles, and the analysis of the recorded sky brigtness is substantially simplified. The zenithal night sky brightness at any given site is a variable whose statistical distribution is determined by the time dependence of the artificial light emissions, the changing atmospheric conditions and the particular configuration of the sky at the time of measuring. A systematic study of the propagation of artificial light through the atmosphere requires the use of large datasets, extended in time and space, to cope with the unavoidable random and deterministic variations of these factors. A permanent network of buoy-based stations in some marine areas could provide the required data. In the meantime, night sky brightness detectors installed on vessels travelling across regular rutes of maritime traffic, as well as on ships in specific measurement expeditions, can provide valuable data for a better understanding of the artificial light propagation through the atmosphere.

**Acknowledgments**

This work was developed within the framework of the Spanish Network for Light Pollution Studies, REECL (AYA2015-71542-REDT). The SQM data of the Barcelona metropolitan area and the Centre d'Observació de l'Univers (COU), provided by Parc Astronòmic Montsec and Consell Comarcal de la Noguera, are gratefully acknowledged. SQM data from Observatori Astronòmic del Montsec were provided by the Direcció General de Qualitat Ambiental i Canvi Climàtic, Departament de Territori i

Sostenibilitat, Generalitat de Catalunya. The REECL SQM detector at Cal Maciarol is operated by Josep Lluis Saltó. We thank the PI Jose M. Baldasano and Michael Sicard for their effort in establishing and maintaining the AERONET Barcelona site. J.Z. acknowledges the support from STARS4ALL H2020-ICT-2015-688135, and S.B. from ED431B 2017/64, Xunta de Galicia/FEDER. Fruitful comments by A. Sánchez de Miguel and C.C.M. Kyba on a draft version of this paper, as well as detailed and insightful comments by two anonymous reviewers were instrumental to improve this work. Any remaining error or omission is of course the sole responsibility of the authors. We express our warmest thanks to Prof. Jordi Torra for his contribution to the initial discussion of the concept of measuring in the sea.

**References**

1. Cinzano P, Falchi F, Elvidge C. The first world atlas of the artificial night sky brightness. Mon. Not. R. Astron. Soc. 2001; 328:689–707.

2. Falchi F, Cinzano P, Duriscoe D, Kyba CCM, Elvidge CD, Baugh K, Portnov BA, Rybnikova NA, Furgoni R. The new world atlas of artificial night sky brightness. Sci. Adv. 2016; 2:e1600377. (doi: 10.1126/sciadv.1600377)

3. Kyba CCM. et al. Worldwide variations in artificial skyglow. Sci. Rep. 2015;5: 8409. (doi:10.1038/srep08409).

4. Bará S, Anthropogenic disruption of the night sky darkness in urban and rural areas, Royal Society Open Science 3:160541 (2016). (doi: 10.1098/rsos.160541).

5. Garstang RH. Model for artificial night-sky illumination. Publ. Astron. Soc. Pac. 1986;98:364-375.

6. Cinzano P, Elvidge CD. Night sky brightness at sites from DMSP-OLS satellite measurements. Mon. Not. R. Astron. Soc. 2004;353:1107–1116 (doi:10.1111/j.1365-2966.2004.08132.x)

7. Kocifaj M. Light-pollution model for cloudy and cloudless night skies with ground-based light sources. Applied Optics 2007;46:3013-3022.

8. Cinzano P, Falchi F. The propagation of light pollution in the atmosphere. Mon. Not. R. Astron. Soc. 2012;427:3337–3357. (doi:10.1111/j.1365-2966.2012.21884.x)

9. Aubé M. Physical behaviour of anthropogenic light propagation into the nocturnal environment. Phil. Trans. R. Soc. B 2015;370:20140117. (doi:10.1098/rstb.2014.0117)

10. Kocifaj M. A review of the theoretical and numerical approaches to modeling skyglow: Iterative approach to RTE, MSOS, and two-stream approximation. Journal of Quantitative Spectroscopy & Radiative Transfer 2016;181:2–10.

11. Longcore T, Rich C. Ecological light pollution. Frontiers in Ecology and the Environment 2004;2:191-198.

12. Rich C, Longcore T (eds). Ecological consequences of artificial night lighting. Washington, D.C.: Island Press (2006).

13. Navara KJ, Nelson RJ. The dark side of light at night: physiological, epidemiological, and ecological consequences. J. Pineal Res. 2007;43:215-224.

14. Hölker F, Wolter C, Perkin EK, Tockner K. Light pollution as a biodiversity threat. Trends in Ecology and Evolution 2010;25:681-682.

15. Gaston KJ, Bennie J, Davies TW, Hopkins J. The ecological impacts of nighttime light pollution: a mechanistic appraisal. Biological Reviews 2013;88:912–927.

16. Gaston KJ, Duffy JP, Gaston S, Bennie J, Davies TW. Human alteration of natural light cycles: causes and ecological consequences. Oecologia 2014;176:917–931.

17. Hölker F, Moss T, Griefahn B, Kloas W, Voigt CC, Henckel D, Hänel A, Kappeler PM, Völker S, Schwope A, Franke S, Uhrlandt D, Fischer J, Klenke R, Wolter C, Tockner K. The dark side of light: a transdisciplinary research agenda for light pollution policy. Ecology and Society 2010;1(4):13. (www.ecologyandsociety.org/vol15/iss4/art13/).

18. Davies TW, Duffy JP, Bennie J, Gaston KJ. The nature, extent, and ecological implications of marine light pollution. Front Ecol Environ 2014; 12(6): 347–355, (doi:10.1890/130281).

19. Davies TW, Coleman M, Griffith KM, Jenkins SR. Night-time lighting alters the composition of marine epifaunal communities. Biol. Lett. 2015;11:20150080. (doi:/10.1098/rsbl.2015.0080).

20. Davies TW, Duffy JP, Bennie J, Gaston KJ. Stemming the Tide of Light Pollution Encroaching into Marine Protected Areas Conservation Letters 2016;9(3):164–171. (doi: 10.1111/conl.12191).

21. Rodriguez A, Moffett J, Revolt A, Wasiak P, McIntosh RR, Sutherland D, Renwick L, Dann P, Chiaradia A. Light Pollution and Seabird Fledglings: Targeting Efforts in Rescue Programs. The Journal of Wildlife Management 2017. (doi: 10.1002/jwmg.21237).

22. Kyba CCM, Ruhtz T, Fischer J and Hölker F, Red is the new black: how the colour of urban skyglow varies with cloud cover. Mon. Not. R. Astron. Soc. 2012;425:701–708. (doi:10.1111/j.1365-2966.2012.21559.x).

23. Solano Lamphar HA, Kocifaj M. Urban night-sky luminance due to different cloud types: A numerical experiment. Lighting Res.& Technol. 2016; 48:1017-1033. (doi 10.1177/1477153515597732).

24. Ribas SJ, Torra J, Figueras F, Paricio S, Canal-Domingo R. How Clouds are Amplifying (or not) the Effects of ALAN. International Journal of Sustainable Lighting 2016;35:32-39. (doi:10.22644/ijsl.2016.35.1.032).

25. Aubé M, Kocifaj M, Zamorano J, Solano Lamphar HA, Sánchez de Miguel A. The spectral amplification effect of clouds to the night sky radiance in Madrid. Journal of Quantitative Spectroscopy & Radiative Transfer 2016;181:11–23.

26. Kyba CCM, Mohar A, Posch T. How bright is moonlight? Astronomy & Geophysics 2017;58(1):1.31-31.32

27. Garcia-Gil M. Contaminació lumínica: fenomen, efectes i abast. Atzavara, L' [on-line], 2017;27:81-87. Last accessed May 2nd, 2017.
 http://www.raco.cat/index.php/Atzavara/article/view/320839

28. Jechow A, Kolláth Z, Lerner A, Hölker F, Hänel A, Shashar N, Kyba CCM. Measuring Light Pollution with Fisheye Lens Imagery from A Moving Boat, A Proof of Concept. arXiv:1703.08484v1 [q-bio.OT] 22 Mar 2017

29. Tamir R, Lerner A, Haspel C, Dubinsky Z, Iluz D. The spectral and spatial distribution of light pollution in the waters of the northern Gulf of Aqaba (Eilat). Sci. Rep. 2017;7:42329. (doi: 10.1038/srep42329).

30. Aubé M, Kocifaj M. Using two light-pollution models to investigate artificial sky radiances at Canary Islands observatories. Mon. Not. R. Astron. Soc. 2012;422:819–830. (doi:10.1111/j.1365-2966.2012.20664.x)

31. Biggs JD, Fouché T, Bilki F, Zadnik MG. Measuring and mapping the night sky brightness of Perth, Western Australia. Mon. Not. R. Astron. Soc. 2012;421: 1450–1464. (doi:10.1111/j.1365-2966.2012.20416.x)

32. Pun CSJ, So CW. Night-sky brightness monitoring in Hong Kong. Environmental monitoring and assessment 2012;184(4):2537-2557.

33. Espey B, McCauley J. Initial irish light pollution measurements and a new sky quality meter-based data logger. Lighting Research and Technology 2014;46(1):67-77.

34. Sánchez de Miguel, A. 2015. Variación espacial, temporal y espectral de la contaminación lumínica y sus fuentes: Metodología y resultados. Ph.D. thesis, Universidad Complutense de Madrid. (doi: 10.13140/RG.2.1.2233.7127)

35. Zamorano J, Sánchez de Miguel A, Ocaña F, Pila-Díez B, Gómez Castaño J, Pascual S, Tapia C, Gallego J, Fernández A, Nievas M. Testing sky brightness models against radial dependency: A dense two dimensional survey around the city of Madrid, Spain. Journal of Quantitative Spectroscopy and Radiative Transfer 2016;181:52–66. (doi: 10.1016/j.jqsrt.2016.02.029)
36. Cinzano P. 2005 Night Sky Photometry with Sky Quality Meter. Internal Report No.9, v.1.4. Istituto di Scienza e Tecnologia dell'Inquinamento Luminoso (ISTIL).
37. Pravettoni M, Strepparava D, Cereghetti N, Klett S, Andretta M, Steiger M. Indoor calibration of Sky Quality Meters: Linearity, spectral responsivity and uncertainty analysis. Journal of Quantitative Spectroscopy & Radiative Transfer 2016;181:74–86.
38. Bessell MS. UBVRI photometry II: The Cousins VRI system, its temperature and absolute flux calibration, and relevance for two-dimensional photometry. Publ. Astron. Soc. Pac. 1979;91:589–607.
39. CIE, Commision Internationale de l'Éclairage. CIE 1988 2° SpectralLuminous Efficiency Function for Photopic Vision. Vienna: Bureau Central de la CIE. (1990).
40. CENELEC, Degrees of protection provided by enclosures (IP Code), EN 60529 standard, European Committee for Electrotechnical Standardization, 1991.
41. Meeus J. Astronomical Algorithms, 2nd ed. William Bell (Richmond, USA) 1998. p. 115
42. Falchi F, Cinzano P, Duriscoe D, Kyba CCM, Elvidge CD, Baugh K, Portnov B, Rybnikova NA, Furgoni R. Supplement to: The New World Atlas of Artificial Night Sky Brightness. GFZ Data Services (2016). (doi:10.5880/GFZ.1.4.2016.001)
43. Duriscoe DM. Measuring Anthropogenic Sky Glow Using a Natural Sky Brightness Model. Publications of the Astronomical Society of the Pacific, 2013; 125(933):1370-1382. (http://www.jstor.org/stable/10.1086/673888).
44. http://guaix.fis.ucm.es/splpr/SQM-REECL (last accessed April 23rd, 2017).
45. Ribas SJ, 2017. Caracterització de la contaminació lumínica en zones protegides i urbanes, PhD Thesis, Universitat de Barcelona, available at http://www.tdx.cat/handle/10803/396095

46. Cinzano P, Falchi F, Elvidge CD. Naked-eye star visibility and limiting magnitude mapped from DMSP-OLS satellite data. Mon. Not. R. Astron. Soc. 2001;323:34–46.
47. https://aeronet.gsfc.nasa.gov (last accessed May 6th, 2017).
48. Ribas SJ, Aubé M, Bará S, Bouroussis C, Canal-Domingo R, Espey B, Hänel A, Jechow A, Kolláth Z, Marti G, Massana P, Schmidt W, Spoelstra H, Wuchterl G, Zamorano J, Kyba CCM. Report of the 2016 STARS4ALL/LoNNe Intercomparison Campaign, doi: http://doi.org/10.2312/GFZ.1.4.2017.001
49. Bará S. Variations on a classical theme: On the formal relationship between magnitudes per square arcsecond and luminance, International Journal of Sustainable Lighting IJSL, 2017; 104-111.
50. Sánchez de Miguel A, Aubé M, Zamorano J, Kocifaj M, Roby J, Tapia C. Sky Quality Meter measurements in a colour-changing world. Monthly Notices of the Royal Astronomical Society, 2017; 467(3):2966–2979. https://doi.org/10.1093/mnras/stx145